\documentclass[natbib]{svjour3}                     
\smartqed  
\usepackage{graphicx}
\usepackage{sidecap}
\usepackage{natbib}
 \newcommand{\aap}{{Astron. Astrophys.}}
 \newcommand{\apj}{{Astrophys. J.}}
 \newcommand{\apjs}{{Astrophys. J. Suppl.}}
 
 \newcommand{\solphys}{{Solar Phys.}}
 \newcommand{\sci}{{Science}}
 \newcommand{\ssr}{{Space Sci. Rev.}}
 \newcommand{\ptrs}{{Phil. Trans. R. Soc. A}}
 \newcommand{\pasj}{{Publ. Astron. Soc. Jpn.}}
\begin{document}

\title{Standing Slow-Mode Waves in Hot Coronal Loops:}
\subtitle{Observations, Modeling, and Coronal Seismology}

\titlerunning{Standing Slow-Mode Waves in Hot Coronal Loops}        

\author{Tongjiang Wang}

\institute{Tongjiang Wang \at
 Department of Physics, Catholic University of America, 620 Michigan Avenue NE,
    Washington, DC 20064, and NASA Goddard Space Flight Center, Code 671, Greenbelt, MD 20771, USA\\ 
              \email{tongjiang.wang@nasa.gov}           
}

\date{received: date / Accepted: date}

\maketitle

\begin{abstract}
Strongly damped Doppler shift oscillations are observed frequently associated with  
flarelike events in hot coronal loops. In this paper, a review of the observed properties and
the theoretical modeling is presented. Statistical measurements of physical parameters 
(period, decay time, and amplitude) have been obtained based on a large number of events observed
by SOHO/SUMER and Yohkoh/BCS. Several pieces of evidence are found to support their interpretation
in terms of the fundamental standing longitudinal slow mode. The high excitation rate of these
oscillations in small- or micro-flares suggest that the slow mode waves are a natural response of 
the coronal plasma to impulsive heating in closed magnetic structure. The strong damping and 
the rapid excitation of the observed waves are two major aspects of the waves that are poorly 
understood, and are the main subject of theoretical modeling. The slow waves are found mainly damped 
by thermal conduction and viscosity in hot coronal loops. The mode coupling seems to play an important 
role in rapid excitation of the standing slow mode. Several seismology applications such as determination 
of the magnetic field, temperature, and density in coronal loops are demonstrated. Further, some
open issues are discussed.
 
\keywords{Solar activity \and Solar corona \and Coronal Seismology \and Coronal loops }
\end{abstract}

\section{Introduction}
The outermost layer of the Sun's atmosphere, the corona, comprises of high temperature plasmas only visible 
in EUV and soft X-ray emissions, which reveal the highly inhomogeneous loop-like structures, believed to 
outline magnetic loops. Coronal loops are highly dynamic. Various periodic and quasi-periodic oscillations 
in radio, visible, EUV, and soft X-rays have been observed for decades \citep[e.g. reviews by][]{asc03, asc04a, 
wan04, ban07}. Most of these oscillations were interpreted as MHD waves. In particular, temporally and spatially 
resolved fast kink mode transverse oscillations and (both standing and propagating) slow mode longitudinal 
oscillations were credibly observed with TRACE satellite and SOHO spacecraft. These high resolution observations 
have turned {\it MHD coronal seismology}, originally suggested by \citet{uch70,rob84}, from a theoretical
concept into a vibrant research area in solar physics \citep[see reviews by][]{nak05, wan05a, dem05, bal07, rob08}. 
Measurements of wave properties (such as periods, amplitudes and damping times) and loop geometry 
in combination with theoretical modeling have been used to estimate physical parameters of coronal plasma 
such as magnetic field \citep{nak01, asc02, wan07}, energy transport coefficients \citep{nak99, ofm02a}, 
density stratification \citep[e.g.][]{and05, van08a}, and loop temperature \citep{mar09, wan09b}. MHD waves 
are also believed to play an important role in coronal dynamics \citep[e.g.,][]{nak07,asc09} and coronal 
heating \citep[e.g.][]{ofm98, asc04b, nak04a, ofm05, tar09}.

Global fast kink-mode transverse oscillations were first found in cool ($\sim$1~MK) coronal loops 
from TRACE EUV imaging 
observations \citep{asc99,nak99}. They were seen as spatial displacements with periods of 3$-$5 minutes 
and were apparently excited by flares or erupting filaments \citep{sch02,asc02}. In several examples, the 
fundamental mode was found together with the second harmonic (or its first overtone) via time series 
analysis \citep{ver04, van07}. A fundamental vertically polarized kink mode oscillation was first 
identified by \citet{wan04b} using TRACE
observations, and a fundamental horizontally polarized kink mode was recently identified by \citet{ver09}
using STEREO/EUVI. \citet{wan08} found that for many combinations of viewing and loop geometry it is not 
straightforward to distinguish between these two types of kink modes just using time series of images.
The combined fundamental horizontal and vertical kink mode oscillations are also possibly excited as
suggested by TRACE and STEREO/EUVI observations \citep{wan08, asc09}. In addition, the vertical 
kink oscillations were detected in coronal multithreaded loops with ejected cool plasma flows with 
the Hinode/SOT Ca\,{\sc{ii}} H line filter \citep{ofm08}. The global fast kink mode oscillations were also
observed with a 2D Doppler coronagraph working in the Fe\,{\sc{xiv}} 5303\AA\ line at the Norikura 
Solar Observatory \citep{hor07}. Various kinds of damping mechanisms
have been proposed to explain the rapid decay of transverse oscillations observed by TRACE, among which
resonant absorption \citep{rud02, van04, goo08}, phase mixing \citep{nak99, ofm02a}, and lateral 
wave leakage \citep{bra05, ver06, sel07a} are believed to be the most possible ones.  
In addition, recently some studies reported the discovery of Alfv\'{e}n waves, such as ubiquitous 
waves in the corona detected in the time series of Doppler images \citep{tom07, tom09}, transverse waves
in threadlike structure of solar prominences \citep{oka07, nin09}, and the transverse oscillation of 
spicules in the chromosphere \citep{dep07, he09a, he09b}. However, \citet{van08b} suggested that these waves are 
more likely fast kink waves rather than the true (incompressible) Alfv\'{e}n mode that is torsional in
cylindrical plasma structure and hard to resolve by present instruments. These kink waves may be driven 
by solar $p$-modes \citep{dep05, hin08, tom09}. Whether they decay during the travel has not been 
investigated in detail. 

Propagating intensity disturbances were first detected in polar plumes with UVCS and EIT onboard SOHO
\citep{ofm97, def98}, and were interpreted as slow magnetoacoustic waves by \citet{ofm99, ofm00}. Their
propagation, viscous dissipation, and refraction were studied by \citet{ofm00} using  2.5D MHD
model. Subsequently, similar propagating disturbances along coronal loops were observed with SOHO/EIT 
\citep{ber99} and TRACE \citep[e.g.][]{nig99, dem00}. Their amplitudes are 3$-$4\% and the propagating speed
is on the order of 100 km s$^{-1}$ \citep{dem02, mce06a}. These upwardly propagating disturbances were
also interpreted as the slow magnetoacoustic wave \citep{nak00}. \citet{dem02} found that loops 
situated above sunspot umbras show oscillations close to 3 min, whereas non-sunspot loops 
(above plage regions) show oscillations close to 5 min. This close link between observed 
periodicity and location suggests that the waves are driven by the underlying $p$-modes
oscillations \citep{dep05, dem07, hin08}. Direct observational evidence for these
waves propagating upwards along the magnetic field through the chromosphere and transition region into
the lower corona have been found in both the cases of 3 minute oscillations \citep[e.g.][]{bry99a, bry99b,
osh02, mar06} and 5 minute oscillations \citep{mar03, osh06, wan09a}. However, \citet{wan09b} recently
reported on observations of 12 and 25 min harmonics of similar propagating waves in coronal loops
using Hinode/EIS observations. Since their periods are much longer
than the cutoff period at the chromosphere and transition region, the origin of these waves are hardly
explained by the wave leakage of the photospheric $p$-modes. In addition, \citet{wan09b} determined the plasma
temperature near the footpoint region of loops using simultaneous measurements of intensity and Doppler shift
oscillations. \citet{mar09} obtained the temperature of the same order from observation of 3D propagation
of similar waves with STEREO/EUVI. The rapid damping of these waves are mainly due to thermal conduction
\citep{dem03, kli04}. For a full review on propagating slow magnetoacoustic waves in coronal loops,
readers can refer to \citet{dem06, dem09}.  However, it should be noted that some recent studies 
argued that these propagating disturbances in fanlike loops and polar plumes observed with TRACE and EIT 
may not be the waves but likely quasi-periodically triggered upflows by episodic heating events based on  
the Hinode/EIS observations \citep{sak07, har08, mci09, mci10, he10, de10}. The detailed discussions are beyond 
the scope of this review.

Not only have the flare-excited transverse kink-mode oscillations but also the flare-excited longitudinal 
slow-mode oscillations have been observed in coronal loops. This review will focused on 
a discussion of the standing slow-mode waves in hot ($>$6 MK) loops observed by the SUMER spectrometer 
on SOHO, including the physical properties, their excitation and damping mechanisms, and
some examples in coronal seismology. The SUMER Doppler shift oscillations will be compared with 
those observed by BCS onboard Yohkoh and recent results from EIS onboard Hinode. For the important 
role of standing slow mode waves in diagnostics of coronal heating, see a recent 
review by \citet{tar09}.  

\begin{figure}
\includegraphics[width=0.6\textwidth]{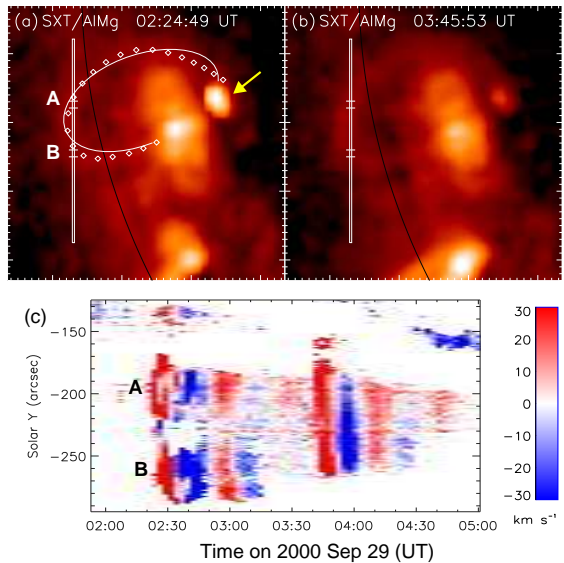}
\centering
\caption{Observations of an oscillating hot coronal loop. (a) and (b) The Yohkoh/SXT images, showing that
the Doppler-shift oscillation events may be triggered by microflares near a footpoint of the loop (marked
with an arrow). (c) Doppler-shift time series in Fe\,{\sc{xix}} observed with SOHO/SUMER in the sit-and-stare mode.
Adapted from \citet{wan03b}.}
\label{fglop} 
\end{figure}

\section{Properties of Observed Oscillations}

\subsection{Mode identification}
The presence of slow mode oscillations in coronal loops was inferred  by long-periodic 
(10$-$30 min) flux pulsations often seen in radio and X-ray wavelengths 
\citep{sve94, asc03}.  Their first direct evidence was provided by \citet{wan02} using 
a high-resolution imaging spectrometer, SUMER, onboard SOHO. Strongly damped Doppler-shift oscillations 
show up in hot flare lines, Fe\,{\sc{xix}} and Fe\,{\sc{xxi}}, with formation temperature greater
than 6 MK \citep{kli02, wan03a, wan03b}. Figure~\ref{fglop} shows an example, in which the coherent Doppler shift
oscillations coincide with the regions where the slit crosses a hot loop seen in X-rays. \citet{wan02} analyzed
two recurring events with coordinated SUMER and Yohkoh/SXT observations. From the measured loop length
($L\approx$140 Mm) and oscillation periods ($P$=14$-$18 min), they estimated the phase speed for the fundamental mode, 
$c_t=2L/P$=240$-$380 km~s$^{-1}$, which are close to the sound speed at about 6 MK, thus suggesting an 
interpretation in terms of the standing fundamental slow mode. For seven oscillation 
events, \citet{wan07} derived the temperature of loops in the range of 6$-$7 MK and the electron density on 
the order of $10^9-10^{10}$ cm$^{-3}$ from the Yohkoh/SXT data. They estimated the period of the fundamental mode
as $P\sim2L/c_s$ where $c_s$ is the sound speed, and found the ratio of the predicted period to the measured period 
in the range  0.75$-$0.94, strongly supporting that the detected loop oscillations all 
belong to the fundamental slow mode. \citet{wan03b} measured physical parameters for 54 oscillation cases 
(Table~\ref{tbpar}) and found the oscillation periods in the range of 7$-$31 min 
with a mean of 17.6$\pm$5.4 min, which is distinctly larger than those (on average about 5 
min) for the TRACE transverse oscillations \citep{asc02}.

\begin{table}
\caption{Comparisons between physical properties of oscillations observed by
SUMER \citep{wan03b}, Yohkoh/BCS \citep{mari06}, and TRACE \citep{asc02, ofm02a}.}
\label{tbpar}
\begin{tabular}{lccc}
\hline\noalign{\smallskip}
            Parameter                  &     SUMER &  BCS  &   TRACE   \\
\noalign{\smallskip}\hline\noalign{\smallskip}
 Oscillation period (minutes) ........  &  17.6$\pm$5.4 &  5.5$\pm$2.7 &  5.4$\pm$2.3 \\   
 Decay time (minutes).................... &  14.6$\pm$7.0 &  5.0$\pm$2.5 &  9.7$\pm$6.4 \\   
 Amplitude (km s$^{-1}$)......................& 98$\pm$75 & 17.$\pm$17.  & 42$\pm$53 \\      
 Displacement (Mm)......................& 12.5$\pm$9.9 & 1.1$\pm$1.7 & 2.2$\pm$2.8 \\ 
 Decay time to period ratio...........& 0.85$\pm$0.35 & 1.05$\pm$0.63 & 1.8$\pm$0.8 \\
\noalign{\smallskip}\hline
\end{tabular}
\end{table}

\begin{figure}
\includegraphics[width=0.95\textwidth]{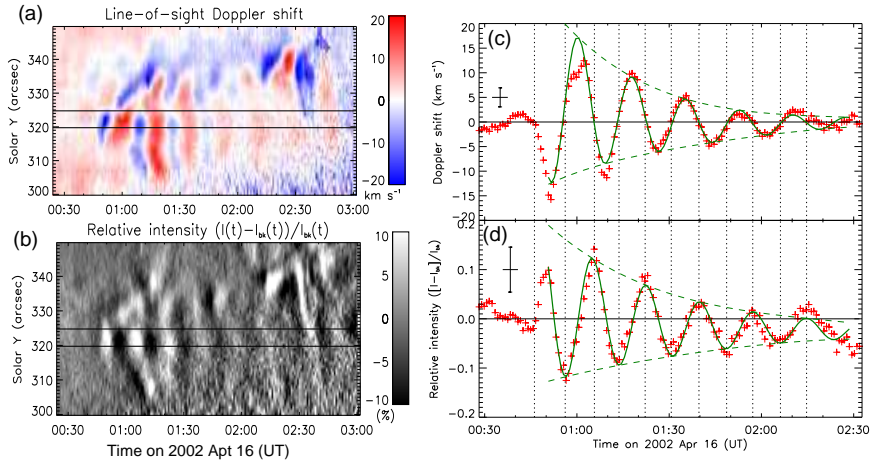}
\centering
\caption{Observations of standing slow-mode waves in a flare loop with SUMER. (a) Time series of Doppler
shift in Fe\,{\sc{xix}} along the slit. (b) Time series of the relative intensity of Fe\,{\sc{xix}}, normalized to the
background emission. (c) and (d) Time profiles of average Doppler shift and relative intensity over a region
along the slit marked in (a) and (b). The solid curves are the best fits with a damped sine function.
The dashed curves are the fitted envelope. }
\label{fgosc} 
\end{figure}

The interpretation of the SUMER oscillations
as the standing slow mode is further supported by the evidence that the associated intensity 
variations show roughly a quarter-period phase delay to the Doppler signal in some cases \citep{wan03a,wan03b}. 
Figure~\ref{fgosc} shows one of the clearest examples. A quarter-period phase lag between velocity and 
intensity disturbances is a characteristic of the standing compressive wave. In contrast, the propagating wave
shows an in-phase relationship \citep{sak02}. 

\begin{figure}
\includegraphics[width=0.8\textwidth]{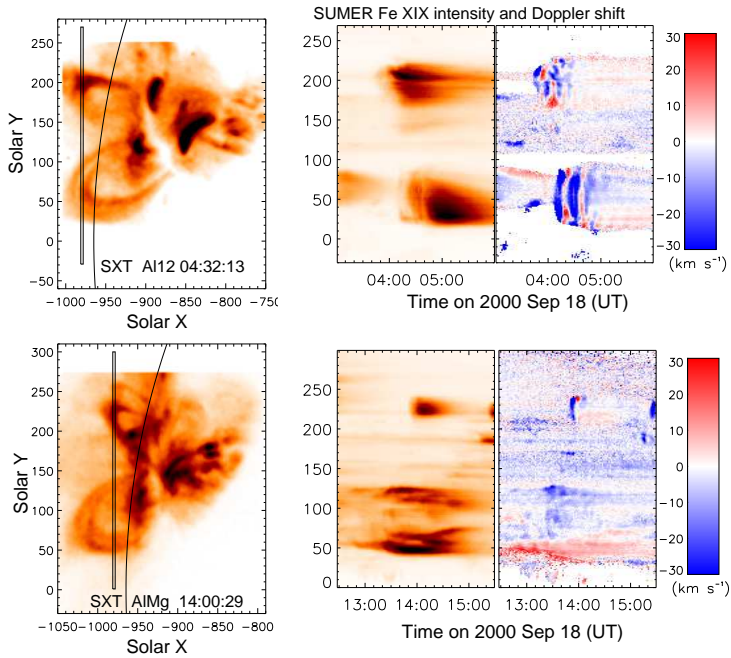}
\centering
\caption{Comparison of two brightening events observed with Yohkoh/SXT ({\it left panels}) and SOHO/SUMER
({\it middle and right panels}). {\it Upper panels}: The case with the slit located at the loop apex.
{\it Bottom panels}: the case with the slit located at the loop legs. 
Adapted from \citet{wan07}. (Reproduced by permission of the AAS)}
\label{fgcmp} 
\end{figure}

In addition, the SUMER oscillations  in
the fundamental mode are also evidenced by the spatial distribution of Doppler shift amplitudes along the loop.
Figure~\ref{fgcmp} shows a comparison of two brightening events observed by SUMER in the Fe\,{\sc{xix}} line above 
a limb active region. The Doppler shift oscillations are clearly seen in the case when the slit is located at
the loop top, whereas no oscillations but plasma flows are seen in the other case when the slit is located
across the loop legs. The fact is consistent with the presence of an anti-node at the loop top 
in velocity perturbation for the fundamental slow mode. 

The detection of similar Doppler-shift oscillations during solar flares in emission lines of S\,{\sc{xv}} and 
Ca\,{\sc{xix}} with BCS on Yohkoh were reported by \citet{mari05}. From measurements of 20 flares,
\citet{mari06} obtained average oscillation periods of 5.5$\pm$2.7 min and decay times of 5.0$\pm$2.5 min . 
Comparison of the averages with the TRACE and SUMER results show that the BCS values are closer to those
from TRACE than those from SUMER (Table~\ref{tbpar}). Since the BCS has no imaging capability, it is 
not possible to determine from the Doppler shift data alone whether the observed oscillations are the fast kink 
mode or slow mode. However, the existence of intensity fluctuations shifted by a quarter-period phase in some
of flares argues strongly that all of the flare Doppler shift oscillations observed with BCS are slow-mode
standing waves. In addition, this interpretation is also supported by the agreement between the
loop length measured from SXT and the one derived from the observed wave period based on an assumption
of the fundamental slow mode. 

Therefore, it is most likely that the SUMER and BCS have observed the same phenomenon. 
Their difference in the period coverages may be caused by selective effects of the instruments. 
The temperature of hot oscillating loops observed by SUMER varies in the range about 6$-$8 MK as measured 
from the SXT data \citep{wan07}, whereas during the early phase of flares when the BCS oscillations 
are observed, the temperature measured in the S\,{\sc{xv}} and Ca\,{\sc{xix}} channels are high upto 
12$-$14 MK \citep{mari06}. Thus, the BCS tends to catch hotter loops which implies the higher 
phase speed, so the shorter wave period. The other reason (probably the major one) is that the BCS data 
often only extend for 20 min or less so that they prefer to detect the Doppler shift oscillations with the 
period less than 10 min \citep{mari06}, while the SUMER data are mostly often acquired 
50$^{''}$$-$100$^{''}$ off the limb, so prefer to detect oscillations of large loops with longer 
periods \citep{wan03b}. 

For both the SUMER and BCS Doppler-shift oscillations only a few of cases are found to exhibit the 
intensity fluctuations with phase lags that are consistent with that expected for a standing slow mode wave. 
This may be explained by the spatial distribution of the amplitude in density perturbation along a loop 
for the fundamental mode, i.e. a node at the loop top and two antiphase
antinodes at the footpoints. In the SUMER case, it is difficult to see intensity oscillations since the slit
is located in most cases at the loop top \citep{wan07}, while in the BCS case  one expects to see intensity 
oscillations only for the partially occulted flares since the BCS images the entire loop.  

\subsection{Triggers}
\label{scttrg}
Several pieces of evidence suggest that the SUMER Doppler shift oscillations are triggered by small (or
micro-) flares near one footpoint of a coronal loop \citep{wan03b}. Firstly, all the SUMER oscillation events
are only seen in hot flare lines of $T$$>$6 MK. Impulsive intensity time profiles and initial large line
broadenings indicate the excitation of oscillations by impulsive heating. Of 27 SUMER flarelike events, \citet{wan03b} 
found that 4 were associated with the GOES C-class flares and 2 were the M-class flares. The others likely
correspond to microflares with an X-ray flux below the detection threshold of GOES. Secondly, the SUMER oscillation
events are characteristic of a high recurrence rate. More than half of the 27 events belong to recurring events 
(with a rate of 2$-$3 times within a couple of hours), which happen at the same place and manifest identical 
periods and initial Doppler shifts of the same sign (and not associated with CMEs). Thirdly, the initiation
of some events is found associated with the X-ray brightening of a footpoint of the oscillating loop
seen with Yohkoh/SXT and RHESSI. A survey of the SUMER oscillation events with RHESSI data have revealed 
that a dozen of the events were triggered by small flares with a hard X-ray source near footpoints of 
coronal loops (Wang et al. 2010, in preparation). Fourthly, the initiation of the SUMER events is found
associated with a high-speed hot transient flow in coronal loops. \citet{wan05b} examined the evolution of
Fe\,{\sc{xix}} and Fe\,{\sc{xxi}} line profiles in the initial phase for 54 oscillations, and found 
that the nearly half show the presence of two spectral components. The shifted component reaches 
a maximum Doppler velocity on the order of 100$-$300 km~s$^{-1}$. This feature indicates that the initial 
Doppler shifts are most likely caused by a pulse of hot plasma flowing along the loop.

\begin{figure}
\includegraphics[width=0.8\textwidth]{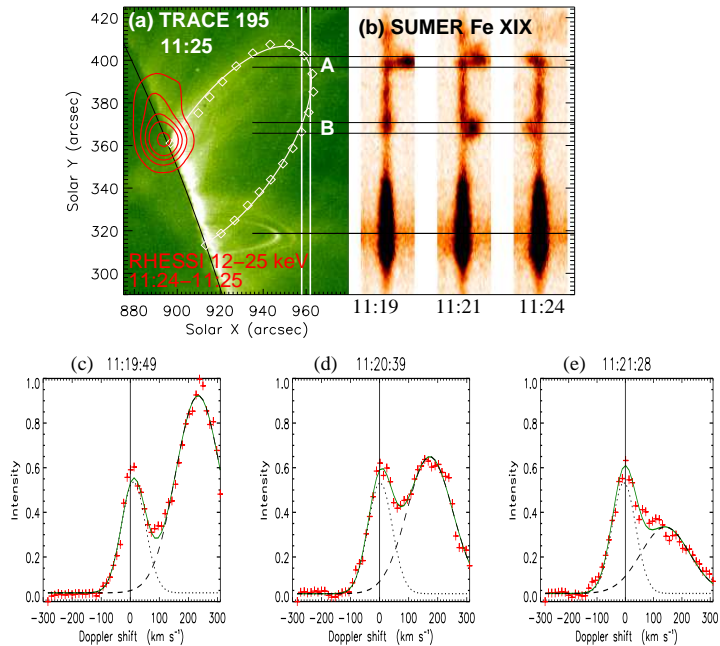}
\centering
\caption{Observations of a hot loop oscillation event at the initial phase. (a) TRACE 195 \AA\ image superimposed
with RHESSI CLEAN image contours at 12-25 keV. (b) The SUMER Fe\,{\sc{xix}} spectra along the slit. (c)-(e) Evolution of
the Fe\,{\sc{xix}} line profiles for position A marked in (b). The dotted and dashed curves are the fitted Gaussian
functions to the two components: one representing the undisturbed loop emission, the other the injected high-speed 
flow emission. Adapted from \citet{wan05b}.}
\label{fgflw} 
\end{figure}

The features described above suggest a scenario for the trigger of SUMER oscillation
events as follows. Small flares are triggered by a sudden energy release due to interactions (local magnetic 
reconnection) between two magnetic flux systems (e.g., a pre-existing large loop and a small twisted emerging 
loop), which produce so-called ``confined flares'' with no substantial change of the magnetic structure 
(and in particular no opening of the closed field system) \citep{hey77, dem97, kar98, wan99}. The observed initial 
high-speed upflow is consistent with the impulsive chromospheric evaporation produced in flares 
\citep[e.g.][]{fis85, dos05, mil09}, while the standing slow mode waves in flare loops are excited by a shock front 
or blast compression disturbance preceding the initial hot plasma flow. 

Figure~\ref{fglop} shows a good example supporting this picture. In this example, two recurring oscillation 
events are associated with a footpoint brightening and both events begin with the redshifts \citep{wan03b},
suggesting that the events are triggered by an impulsive energy release near the footpoint of the
oscillatory loop, while the initial redshifts are produced by hot evaporation (or reconnection) flows.
Figure~\ref{fgflw} demonstrates another clear example. This event initiates as two 
brightenings separated by a time lag and located near positions where the slit intersects a coronal loop. 
RHESSI observations show a hard X-ray source located at a footpoint of the loop, suggesting that
a flare near the footpoint produces the high-speed hot flow and excites the slow mode oscillations
in the loop. The evolution of spectral profiles shows that the oscillation set up immediately after 
the decay of the initial hot flow with a speed more than 200 km~s$^{-1}$ \citep{wan05b}.  

The slow-mode oscillations observed by SUMER and BCS both appear to show a high excitation rate by small 
flares. \citet{wan06} studied the frequency distribution of thermal energy for small flarelike 
brightenings observed with SUMER in Fe\,{\sc{xix}} above active regions on the limb. These brightenings
have lifetimes ranging from 5 to 150 min with an average of about 25 min, and their estimated thermal energy
from 10$^{26}$ to 10$^{29}$ erg with a power law index of 1.7 to 1.8. The features are well consistent with
soft X-ray active region transient brightenings observed by Yohkoh/SXT \citep{shi95}, suggesting that these 
Fe\,{\sc{xix}} brightenings are the coronal parts of loops heated to more than 6 MK by soft X-ray microflares.  
For nearly 300 events identified visually, 40\% of them are associated with Doppler shift oscillations. 
Those events without showing evident oscillations may be ascribed to their observed location not close to 
the loop top (e.g., see Fig.~\ref{fgcmp}). \citet{mari06} examined 103 flares observed by Yohkoh/BCS
and found 38 showing oscillatory behavior in the measured Doppler shifts. Particularly, 20 of them with
well-defined oscillations are mostly located at the limb. This location preference may be caused 
by the projection effect. As the longitudinal velocity perturbation (parallel to the field) has 
the maximum amplitude at the loop top
for the fundamental mode, it tends to produce smaller Doppler-shift signals when a flare occurs in the loops 
located close to the disk center. This implies that the BCS Doppler-shift oscillations may be excited 
indeed more frequently than the observed. In addition, as the SUMER oscillations the BCS oscillations also 
prefer to be triggered by small flares. Of 20 events studied by \citet{mari06}, 75\% are C-class flares 
(the rest are M-class). 

In summary, the above discussions suggest that standing slow mode oscillations 
are a nature of the confined flares, i.e. they are commonly excited by impulsive energy release in 
non-eruptive closed magnetic structure. Their property of the high excitation and occurrence 
rates make it invaluable for coronal seismology (see Sect.~\ref{sctss}).

\section{Theoretical Modeling}

\subsection{Excitation}
For modeling the excitation of slow mode standing waves in hot loops observed by SUMER, 
\citet{wan05b} suggested that consideration of the following observed properties are important.\\
1. The standing slow-mode waves in flaring loops are set up within one wave period. \\
2. The oscillations are the fundamental mode which appears to be triggered by impulsive asymmetric 
   heating at one footpoint of the loop. The heating time is less than about half of the wave period,
   as implied by the duration of initial line width broadenings.\\
3. The loop plasma is impulsively heated to above 6 MK in the initial phase of events, 
   and then gradually cools back to the initial temperature of about 2$-$3 MK. 

\begin{figure}
\includegraphics[width=1.0\textwidth]{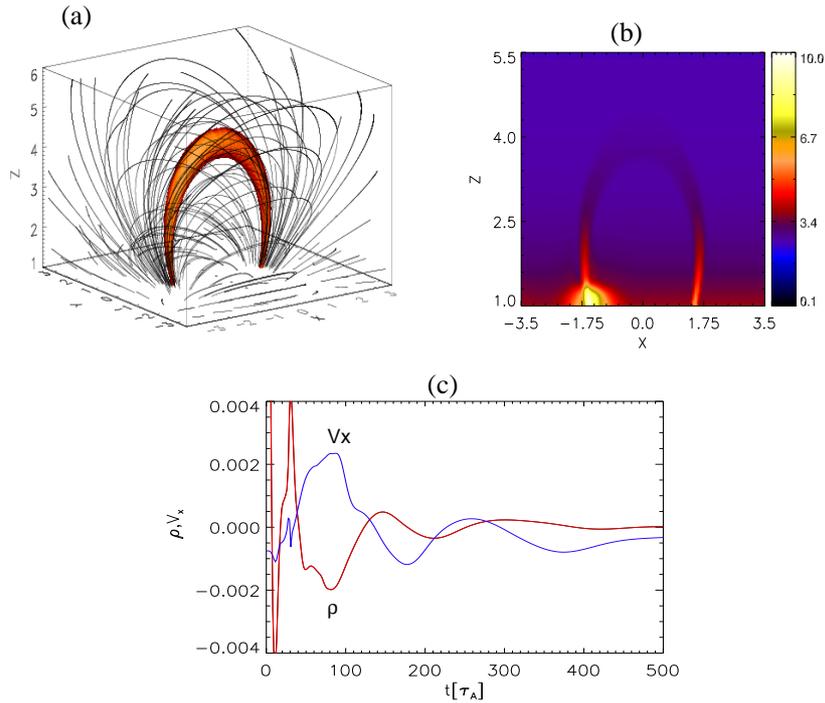}
\centering
\caption{3D MHD simulation of slow standing waves in a curved active region loop.
(a) Initial 3D magnetic configuration of the model AR with a high density loop. (b) Mass density profiles in x-z plane 
(for y =0) at the beginning of simulation, when the pulse appears. (c) Time-signatures of the mass density and 
x-component of velocity at the loop's apex.  Adapted from \citet{sel09}.}
\label{fgsim} 
\end{figure}

The excitation of standing slow mode waves in hot coronal loops has been theoretically studied by many
authors using 1D, 2D and 3D MHD models. In the 1D loop model, the magnetic field plays only a role to guide
the wave. The advantage is that highly complex field-aligned energy transport processes can be accurately
described using a full energy equation. \citet{sel05} showed that pressure pulses launched close to a 
footpoint excite the fundamental mode, but forming the standing waves needs a dozen or so wave periods. 
\citet{tar05} analytically studied the excitation condition for the fundamental mode by the footpoint 
heating and found that to rapidly set up a standing wave in a single period it needs the duration of the
heating pulse approximately matches the period of the oscillations. Under this assumption, \citet{tar07}
studied a SUMER oscillation event using the forward modeling method. They estimated the duration,
temporal behavior, and total heat input of the microflare and recovered the time-distance profile of the 
heating rate along the loop. Based on simulations of the similar model, \citet{tar08} predicted footprints 
and observables of standing and propagating acoustic waves in Hinode/EIS data. \citet{pat06} also
generated the synthetic line profile observations based on simulations of the loop oscillation produced
by impulsive heating.

However, the duration of the heating pulse required to excite the standing wave in  \citet{tar07} 
is too long to be supported by observations. This issue may come from the simple treatment of an actual 3D 
curved loop to the 1D model. \citet{ogr07} showed that slow waves are excited faster in 2D straight
slab than in a 1D loop due to coupling between the fast and slow magnetoacoustic waves. \citet{sel07b} 
considered a 2D arcade loop model and found that the curved configuration plays an important role in 
efficient excitation of a slow standing wave because of the combined effect of the pulse inside and outside
the loop. The standing wave can be set up within 1.6 wave periods. \citet{ogr09} performed a parametric 
study using the similar 2D model. Recently, \citet{sel09} extended the 2D arcade model to the
3D loop model in a dipolar field. Figure~\ref{fgsim} shows that a velocity pulse covering both the footpoint 
of the loop and surrounding plasma can excite the standing fundamental mode of slow waves within a time
interval comparable to the observations. The key point of the 2D and 3D models different from the 1D model
is that a fast-mode component of the initial pulse is allowed to propagate quickly outside the loop and
excites another slow pulse at the opposite footpoint almost simultaneously as the initial slow pulse
(with respect to the time scale of the slow wave period). This effect may cause a great reduction in the
excitation time of the standing wave. \citet{pas09} also studied impulsively generated oscillations in
a 3D coronal loop, and found that a fundamental slow mode wave is excited when a pressure pulse nearby
impacts on the loop almost parallel to the loop plane, but they did not mention the excitation time.
Simulations of the excitation of standing slow waves in a 2D or 3D loop model by impulsive heating 
are required in the future for developing heating diagnostics of hot coronal loops with MHD waves. 

In addition, \citet{hay08} performed a 3D modeling of the kink instability in a twisted straight coronal
flux tube, and found that the kink instability initially sets up the second harmonic that is then 
converted into two out-of-phase fundamental slow modes in two newly-formed, entwined structures. 
However, no evidence has been found that the SUMER oscillations are triggered by the kink instability.
 
\begin{SCfigure}
\includegraphics[width=0.5\textwidth]{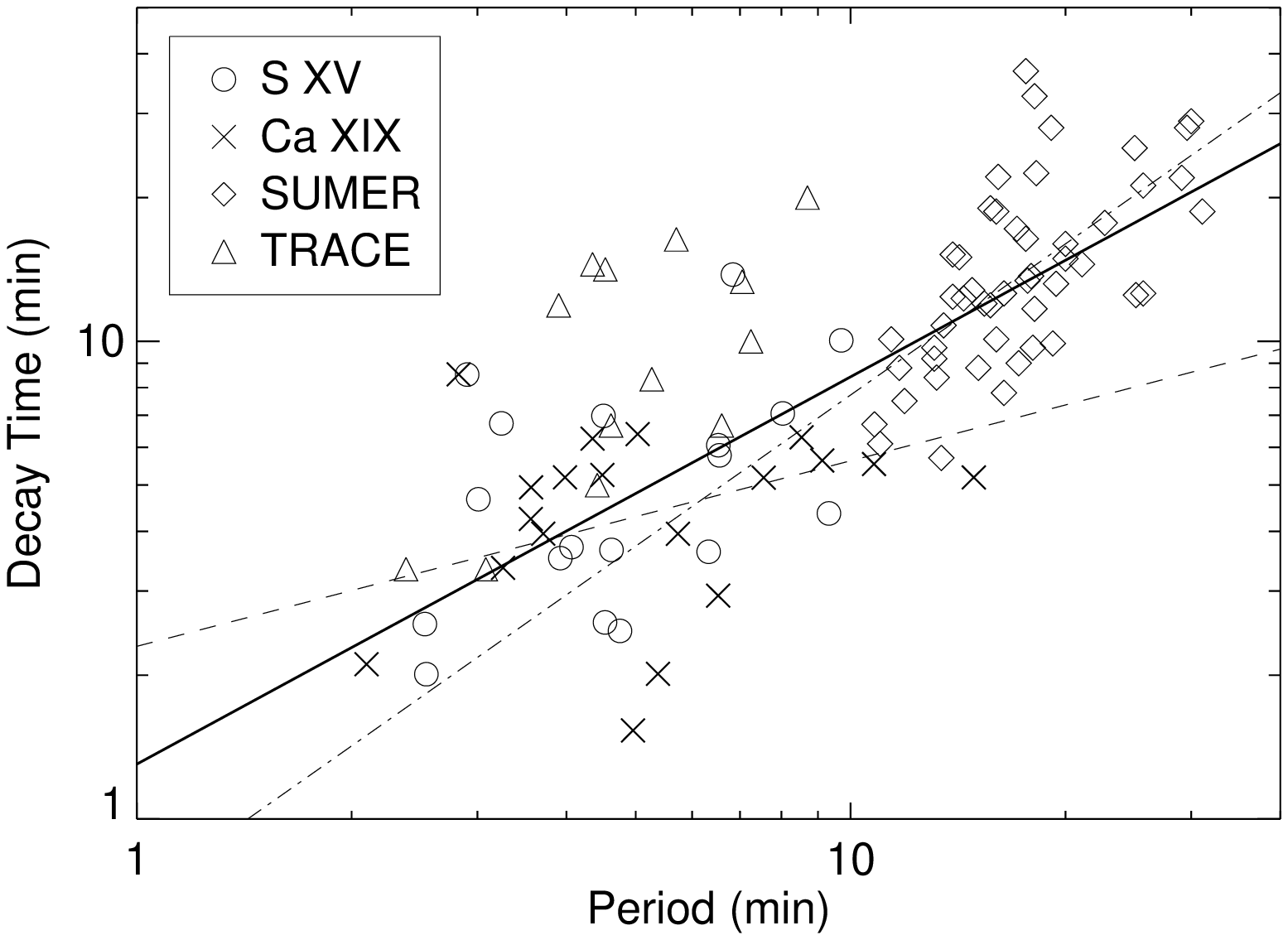}
\caption{Measured decay time vs. period for the events observed with the
BCS reported by \citet{mari06}, the SUMER events reported by \citep{wan03b}, and the TRACE
events reported by \citet{ofm02a}, \citet{wan04b}, and \citet{ver04}. The dashed line shows 
the best-fit to the BCS data, the dot-dashed line shows the fit to the SUMER data, and the solid line 
shows the fit to the combined BCS and SUMER data sets. Adapted from \citet{mari06}. 
(Reproduced by permission of the AAS)}
\label{fgdmp} 
\end{SCfigure}

\subsection{Damping}
The Doppler shift oscillations observed by both SUMER and BCS decay strongly, with a ratio of 
the damping time to the period on the order of 1 (see Table~\ref{tbpar}). Figure~\ref{fgdmp} shows the decay
time of the oscillations plotted against the period for the events observed with SUMER, BCS and TRACE.
The fit to the combined SUMER and BCS data results corresponds to the expression $t_{decay}=1.30P^{0.81}$
\citep{mari06}, which is very close to the fit to the SUMER data alone \citep{wan03b}. 

\citet{pan06} argued that the individual effect of thermal conduction or viscosity is not sufficient to 
explain the observed damping based on the linear analysis applicable for small amplitude waves. \citet{sig07} 
compared the damping of standing slow waves due to various dissipation effects in both the linear and 
nonlinear regimes and obtained the similar conclusion. However,  \citet{ofm02b} showed using nonlinear
1D MHD modeling that thermal conduction is the primary dissipation mechanism of the observed large 
amplitude standing slow-mode waves in hot coronal loops, and the scaling of the dissipation time
with period agrees well with the SUMER oscillations. \citet{men04} showed that the damping times
are reduced by 10$-$20\% compared to homogeneous loops because of enhanced nonlinear viscous dissipation
induced by gravity.  \citet{bra08} studied the
effect of the radiative emission arising from a non-equilibrium ionization on the damping, and found that 
it may reduce the damping timescale by up to $\sim$10\% compared to the equilibrium case. 
\citet{hay08} showed that in the absence of thermal conduction, large-amplitude slow-mode oscillations
are still damped rapidly due to shock dissipation. \citet{ver08} further studied the case in the
presence of thermal conduction, and found that shock dissipation at large amplitudes enhances the damping 
rate by up to 50\% above the rate given by thermal conduction alone. 
\citet{erd08a} studied the damping of standing slow waves in 
nonisothermal, hot, gravitationally stratified coronal loops, and found that the decay time of waves
decreases with the increase of the initial temperature. They also derived a second-order scaling 
polynomial between the damping time and the parameter determining the apex temperature. \citet{ogr07}
found that standing slow waves are attenuated more efficiently in 2D straight slab than in 1D loop due 
to coupling between the fast and slow magnetoacoustic waves. \citet{ogr09} further showed that these waves 
are attenuated more quickly in a curved 2D arcade than in a straight 2D slab 
configuration due to the possible enhanced wave leakage. \citet{sel09} found that the damping of
slow waves in the curved 3D loop is faster than in the curved 2D loop, and explained it in terms of
lateral leakage because the 3D geometry provides additional degrees of freedom for the wave to leak 
out of the loop than 2D models. However, since these 2D and 3D MHD models metioned above do not
include the dissipation terms such as thermal conduction and viscosity in the energy equation, 
it is unclear whether the coupling to the fast wave due to the transverse density inhomogeneity and 
the lateral leakage due to the loop curvature can dominate over the thermal conduction responsible
for the rapid decay of observed standing slow-mode waves. This needs further investigation in
the future.

\section{Applications of Coronal Seismology}
\label{sctss}
\subsection{Determination of coronal magnetic field}
Magnetic field plays a key role in understanding dynamics and heating of solar corona, however, direct 
measurement of the coronal magnetic field remains a very difficult problem. Recent high-resolution
space observations have made MHD coronal seismology a promising tool to measure the coronal field
indirectly. Table~\ref{tbknk} shows many examples for estimates of the coronal field from the fast 
kink-mode oscillations in coronal loops. In contrast, \citet{wan07} determined the coronal field from
the standing slow-mode oscillations in hot coronal loops observed by SUMER.

\begin{figure}
\includegraphics[width=0.9\textwidth]{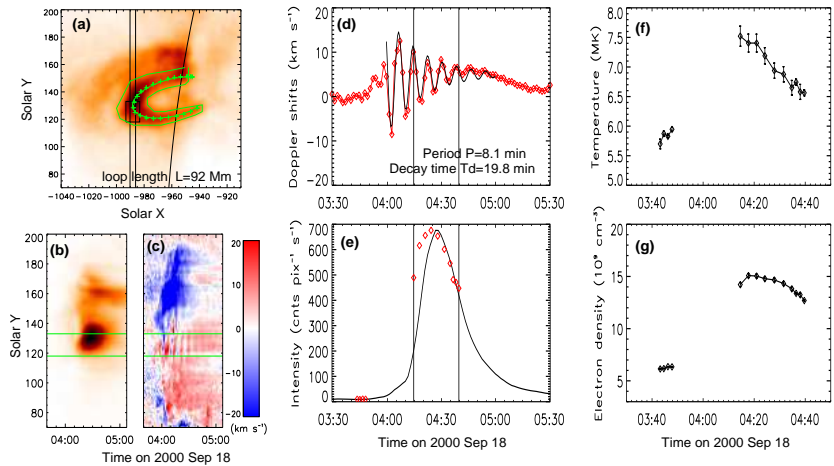}
\centering
\caption{Coordinated observations of a hot loop oscillation event by SOHO/SUMER and Yohkoh/SXT. (a)~The SXT 
image. (b,c)~Line intensity and Doppler shift time series in Fe\,{\sc{xix}} along the slit observed with SUMER.
(d)~Time profile of Doppler shifts averaged over a region marked in (c). The thick solid curve represents the
best fit to a damped sine function. (e) Time profile of line intensities averaged over a region marked in (b). 
The superimposed data ($diamonds$) are the mean intensity of SXT emission averaged over an area at the loop top. 
(f,g)~Temperature and electron density time profiles, calculated for the whole loop with the filter ratio method 
from SXT images. Adapted from \citet{wan07}. (Reproduced by permission of the AAS)}
\label{fgmeg} 
\end{figure}

The relative higher plasma $\beta$ in hot ($>6$ MK) loops implies the stronger coupling between
magnetic perturbation and thermal perturbation than in the typical coronal structure, thus the magnetic field
in hot loops may be deduced from diagnostics of standing slow-mode oscillations. Applying the MHD wave 
theory for a straight magnetic cylindrical model to the coronal loop \citep{edw83}, the oscillation 
period of a standing slow mode wave in its fundamental mode is given by,
\begin{equation}
 P=\frac{2L}{c_t}, \hspace{3mm} c_t=\left(\frac{1}{c_s^2}+\frac{1}{v_A^2}\right)^{-1/2}, \label{eqpslw}
\end{equation}
\noindent
where $L$ is the loop length, $c_s$ is the sound speed, and $v_A$ is the Alfv\'{e}n speed. \citet{rob06}
showed that the effect of gravitational stratification on the oscillation period of slow modes in hot 
SUMER loops is negligible. \citet{erd08a} showed that the effect of longitudinal temperature structure 
on the period is also trivial. Additional source of the uncertainty comes from the thin flux 
tube approximation of real coronal magnetic structure, under which the above expression is derived. 
From equation~(\ref{eqpslw}), the magnetic field can be derived as
\begin{equation}
B=\left(\frac{N_e}{C_1}\right)^{1/2}\left(\frac{P^2}{4L^2}-\frac{1}{C_2T}\right)^{-1/2},  \label{eqmb}
\end{equation}
\noindent
where $C_1$ and $C_2$ are constants, $B$, $N_e$, and $T$ are the field strength, electron density, and
loop temperature, respectively. For 7 Doppler-shift oscillation events simultaneously observed with 
the SUMER and Yohkoh/SXT, \citet{wan07} estimated the magnetic field in the range 21$-$61 G with a mean
of 34$\pm$14 G. The subtraction of background emission leads to a reduction of the estimated values by
9\%$-$35\% (see Table~\ref{tbslw}). The error propagation analysis shows that the uncertainty in the
estimated magnetic field is mainly contributed from uncertainties in temperature and loop length 
measurements which are reversely proportional to the plasma $\beta$. Figure~\ref{fgmeg} shows an example to
demonstrate the measurements of $P$, $L$, $N_e$ and $T$. The period is measured by fitting the Doppler
shift oscillations with a damped sine function. The loop length is measured by fitting the SXT loop with
a simple geometric loop model. The average electron temperature and density are derived using the filter
ratio method. Since $T$ and $N_e$ both evolved with time because of loop heating and cooling, their 
time-averaged values are used to determine $B$. Therefore, to improve the measurement accuracy of $B$, 
not only improved measurements of $T$ and $N_e$ should be conducted, but the effect of their 
temporal variations on the oscillation period also need to consider in the future.

\begin{table}
\caption{Overview of measurements of the coronal magnetic field from observations of fast kink-mode oscillations.}
\label{tbknk}
\begin{tabular}{lllcl}
\hline\noalign{\smallskip}
Reference & Wavelength & Instrument & $B$-field (G) & Loops \\ 
\noalign{\smallskip}\hline\noalign{\smallskip}
\citet{nak01}....... &  171 band & TRACE & 4$-$30 & 2 \\
\citet{asc02}............ &  171 \& 195 bands & TRACE & 3$-$90 & 26 \\
\citet{ver04}................ & 171 band & TRACE & 9$-$46 & 9 \\
\citet{ofm08}................ & Ca II H band & Hinode/SOT & 20$\pm$7 & 1 \\
\citet{van08c}... & Fe XII 195 line & Hinode/EIS & 39$\pm$8 & 1 \\
\citet{erd08b}............ & Fe XII 195 line & Hinode/EIS & 10$\pm$6 & 1 \\
\citet{ver09}................ & 171 band & STEREO/EUVI & 11$\pm$2 & 1 \\
\noalign{\smallskip}\hline
\end{tabular}
\end{table}

\begin{table}
\caption{Average and range of physical parameters for standing slow-mode oscillations in hot coronal loops 
measured from coordinated SUMER and Yohkoh/SXT observations. Adapted from \citet{wan07}.
(Reproduced by permission of the AAS)} 
\label{tbslw}
\begin{tabular}{lccccc}
\hline\noalign{\smallskip}
Parameter  & \multicolumn{2}{c}{Before $I_{bk}^{\mathrm{a}}$ subtraction}& & \multicolumn{2}{c}{After $I_{bk}$ subtraction} \\
  \cline{2-3}  \cline{5-6}\\
   & Average &  Range & & Average & Range \\
\noalign{\smallskip}\hline\noalign{\smallskip}
  Oscillation period $P$ (minutes) .......& 11.2$\pm$3.8  & 8.1$-$18.3 & &\multicolumn{2}{c}{Unchanged} \\
  Loop length $L$ (Mm) ...................... & 116$\pm$44  & 74$-$199  & & \multicolumn{2}{c}{Unchanged} \\
  Temperature $T$ (MK) ....................... & 6.4$\pm$0.3    & 5.9$\pm$7.0& & 6.6$\pm$0.4 & 6.1$-$7.0 \\
  Electron density $N_e$ ($10^9$ cm$^{-3}$) ......& 8.4$\pm$3.6 & 4.3$\pm$14.1& & 7.4$\pm$3.3 & 3.4$-$12.4 \\
  Sound speed $c_s$ (km s$^{-1}$) ..................& 385$\pm$10   & 369$-$402& & 391$\pm$11 & 375$-$402 \\
  Alfv\'{e}n speed $v_A$ (km s$^{-1}$) ..............& 830$\pm$223 & 442$-$1123& & 784$\pm$218 & 438$-$1123 \\
  Plasma $\beta$ .........................................& 0.33$\pm$0.26  & 0.15$-$0.91& & 0.38$\pm$0.27 & 0.15$-$0.94 \\
  Magnetic field $B$ (G) .......................& 34$\pm$14 & 21$-$61& & 31$\pm$14 & 19$-$57\\
\noalign{\smallskip}\hline
\end{tabular}
\begin{list}{}{}
\item[$^{\mathrm{a}}$] 
   $I_{bk}$ is the background emission of a loop.
\end{list}
\end{table}

\subsection{Derivation of temperature evolution}
We may explore temperature evolution of an oscillatory loop based on the relationship between the amplitudes 
of velocity and intensity variations for the slow mode wave. For a propagating wave, the amplitude relationship 
can be easily derived from the linearized continuity equation as
$I^{\prime}/2I\sim \rho^{\prime}/\rho_0= V^{\prime}/c_t$, where $I^{\prime}/I$ and $V^{\prime}$ 
are the relative amplitude of
the intensity and velocity amplitude, respectively, $\rho^{\prime}$ and $\rho_0$ are the density perturbation
and the undisturbed loop density, and $c_t$ is the phase speed. As $\beta\ll{1}$ for a typical (T$\sim$1 MK) 
coronal loop, $c_t\approx{c_s}$.  We thus can determine the sound speed (so the temperature) from 
the ratio between the intensity and velocity amplitudes \citep[e.g.][]{wan09b}. 

\begin{SCfigure}
\centering
\includegraphics[width=0.6\textwidth]{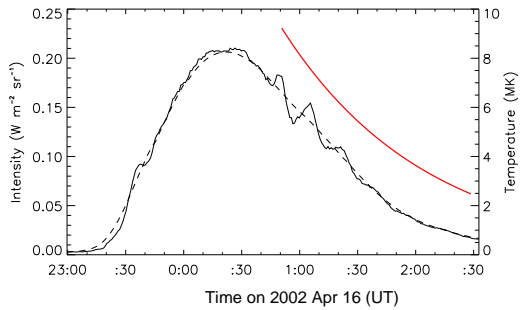}
\caption{Time profile of Fe\,{\sc{xix}} intensity (thin solid curve) observed by SUMER. The dashed curve is 
the 20-min window smoothing of the light curve. The thick solid line shows the cooling curve of loop 
temperature derived from the amplitude ratios between Doppler shift and relative intensity oscillations 
with Eq.~(\ref{eqiv}).}
\label{fgte} 
\end{SCfigure}

For a standing wave the amplitude relationship also depends on the detected position of the oscillations. 
\citet{wan03a} derived it as,
\begin{equation}
  \frac{I^{\prime}}{2I}\sim\frac{\rho^{\prime}}{\rho_0}=\frac{V_x}{c_t}|{\rm tan}(kx)|,
  \label{eqiv}
\end{equation}
where $V_x=V^{\prime}{\rm cos}(kx)$, the amplitude of the detected oscillation at a position  and $x$ is 
its distance to one footpoint along the loop. $k=\pi/L$, the wavenumber for the fundamental mode. 
For an event shown in Figure~\ref{fgosc}, \citet{wan03a} measured $x\approx0.33L$ and the angle of $V_x$ 
to the line-of-sight, $\gamma$=15$^{\circ}$, by fitting the loop geometry with a 3D circular model.
The Doppler shift amplitude ($V_{\|}(t)$) and the relative intensity amplitude 
($I^{\prime}(t)/I(t)$) are measured by fitting the oscillations with an exponentially damped sine function.
We obtain that $V_{\|}(t)=18.3{\rm exp}(-t/33.7)$ km~s$^{-1}$ and $I^{\prime}(t)/I(t)=0.16{\rm exp}(-t/43.6)$, 
where $t$ is in units of minute relative to the start time for fitting. So $V_x(t)=V_{\|}(t)/{\rm cos}\gamma$. 
From Eq.~(\ref{eqpslw}) it follows that
$c_t=c_s/(1+\gamma\beta/2)^{1/2}$. Given $\beta$=0.3, a typical value for the SUMER hot loops \citep{wan07},
then $c_t=0.89c_s$. Finally, using Eq.~(\ref{eqiv}) we obtain the loop temperature, $T(t)=9.2{\rm exp}(-t/74)$ MK.
Figure~\ref{fgte} shows the derived temperature evolution from 00:51 to 02:28 UT, revealing that 
the loop cools from about 9.2 MK to 2.5 MK within a period of 86 min, over which the oscillations are clearly visible.
This result agrees well with that measured in another event, where the SUMER slit is located at a similar height 
on the limb. The cooling time can be estimated as the delay time between the peak of the light curves seen in 
Ca\,{\sc{xiii}} (3.2 MK) and Fe\,{\sc{xix}}, which is about 90 min (see Fig.~\ref{fgco}d).
This example demonstrates that the method proposed above is reasonable. 

The problem of the cooling of hot or post-flare loops was previously studied in both observation and theory
by many authors \citep[e.g.][]{ant78, ant80, cul94, car95, har98, asc01, kam03}.
A well-accepted model by \citet{car95} assumes that conductive cooling initially dominates with radiative 
cooling taking over later on, suggesting that the cooling time of flaring loops can vary in a wide
range depending on parameters such as the initial temperature, the loop length (or height) and the density.
For example, from height$-$time curves of soft X-ray and 
H${\alpha}$ loops, \citet{har98} derived cooling times from 25 to 90 min for initial 
small and dense loops, while more than 100 min for later larger loops in an event using an initial temperature 
of 8 MK, which were found similar to the calculated cooling times using the method of \citet{car95}. 
Our observations of the events discussed above unfortunately do not allow a comparison with the models 
because of the lack of density information although the measured cooling time appears to be
in a reasonable range considering the large loop length ($\sim$200 Mm). The future observations
from the combined Hinode/XRT, EIS and SDO/AIA will provide good opportunities in examining the seismological 
method suggested here in virtue of their abilities to simultaneously measure the temperature, density and 
geometry of the oscillating loop.

\subsection{Diagnosis of plasma density}
\begin{SCfigure}
\centering
\includegraphics[width=0.5\textwidth]{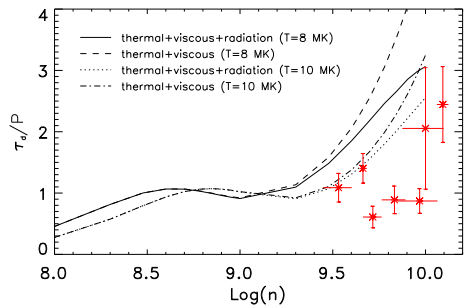}
\caption{Variation of the ratio between the decay time and wave period ($\tau_d/P$) as a function of 
number density ($n$) for a loop length of 400 Mm. The solid and dashed curves correspond to a loop 
temperature of 8 MK, while the dotted and dash-dotted curves correspond to 10 MK 
\citep[Adapted from][]{pan06}. The superimposed data are measurements from the SUMER and SXT observations 
reported by \citet{wan07}.}
\label{fgne} 
\end{SCfigure}

The Doppler shift oscillations observed by SUMER are strongly decayed, with a ratio of the damping
time ($\tau_d$) to the period ($P$) on the order of 1 \citep{wan03b}. There are a few of cases, however, showing 
a peculiar weak decaying with oscillations visible for more than 5 periods and $\tau_d/P\geq2$ \citep{wan03a, wan07}.
\citet{wan03a} suggested that the weak damping may be due to a high plasma density
in some flaring loops. This idea was confirmed by \citet{pan06} based on the linear wave theory.  They found that
for coronal loops with a temperature in the range 6$-$10 MK, the strong-damped ($\tau_d/P$$\sim$1) oscillations 
occur in lower-density (10$^8$$-$10$^9$ cm$^{-3}$) loops, while the weak-damped ($\tau_d/P$$\geq$2) oscillations 
occur in higher-density (10$^9$$-$10$^{10}$ cm$^{-3}$) loops. This result appears to be supported by the coordinated 
SUMER and SXT observations. Figure~\ref{fgne} shows a comparison between the observed result and those predicted 
with and without including the effect of optically thin radiation besides thermal conduction and viscosity.
We find that out of 7 loops, 4 seem to follow the curve predicted in the case
when $T$=10 MK with the effects of thermal conduction and viscosity.  This result suggests that the dependence 
of $\tau_d/P$ on $N_e$ and $T$ may be used to diagnose the density and temperature in oscillatory loops. 
However, given the huge error bars and marginal fit in Fig.~\ref{fgne}, this will at best be an order of 
magnitude estimate in this example. Based on the future observations with Hinode/EIS and SDO/AIA, the validity
of the method suggested here should be examined by comparing the derived plasma electron density 
and temperature with those obtained using the classic methods such as line intensity ratio and differential 
emission measure (DEM) techniques \citep[e.g.][]{bro96, lan03}. 

\section{Discussions}

\subsection{Propagating feature}

\begin{figure}
\includegraphics[width=0.9\textwidth]{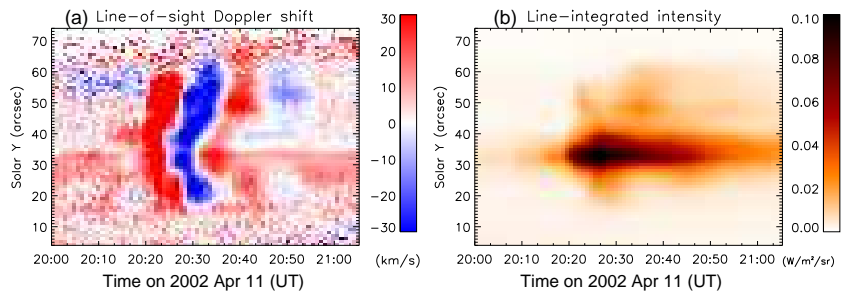}
\centering
\caption{Propagation feature of Doppler shift oscillations revealed by high-cadence (50 s) SUMER
observations. (a) Time series of Doppler shifts in Fe\,{\sc{xix}} along the slit.
(b) Time series of the Fe\,{\sc{xix}} line intensity. }
\label{fgprp} 
\end{figure}

In some cases, high-cadence SUMER observations of Doppler oscillations exhibit the feature of phase propagation
along the slit with speeds in the range 8$-$102 km~s$^{-1}$ and a mean of 43$\pm$25 km~s$^{-1}$. Figure~\ref{fgprp}
shows such an example. The phase delay seems to propagate along the slit from the strong emission region 
to the faint ones. This propagating feature apparently does not agree with the property of the fundamental 
standing waves, which should produce inphase oscillations along the loop. However, this feature is either 
inconsistent with that produced by propagating waves. Simulations by \citet{tar08} showed that the propagating 
slow-mode pulse reflecting back and forth inside the loop exhibits a triangle-like pattern in the time-distance 
plot of Doppler velocity. \citet{hor07} reported simultaneous excitation of transverse fast kink
and longitudinal slow modes in a bundle of coronal loops. In the individual loops, a crossing diagonal pattern 
indicates the propagating slow waves reflecting on both ends of the loop. Instead, \citet{wan03b} proposed 
a scenario to explain the origin of this phase propagation by excitation of standing slow waves in coronal loops
with the (multithreaded) fine structure. If magnetic reconnection triggers thermal energy release at a 
loop's footpoint in a certain thread, the produced gas-pressure disturbance will affect the other threads 
at a slightly later time than the directly involved thread, thus exciting slow waves in those threads with 
phases delayed relative to the slow wave in the thread directly related to the trigger. The 3D MHD loop
model in combination with the forward modeling method is necessary to test this idea.

\subsection{Temperature dependence}
The SUMER Doppler-shift oscillation events are all reported in hot flare lines, Fe\,{\sc{xix}} and 
Fe\,{\sc{xxi}}, with a formation temperature at above 6~MK \citep{wan03b}. Observations of these events at 
cooler lines have not yet been reported. It is not clear whether this implies that the excitation of slow mode 
standing waves requires the heating source with an energy input rate above certain threshold value, or
only because the oscillations excited in loops heated to the lower temperature are too weak to be detected
by SUMER. Recently, small amplitude (2$-$3 km~s$^{-1}$) Doppler-shift oscillations are observed in
Fe\,{\sc{x}}$-$Fe\,{\sc{xv}} lines with Hinode/EIS. Some are consistent with propagating slow magnetoacoustic waves, whose 
excitation appears to be related to the photospheric drivers \citep{wan09a,wan09b, mari10}. While some others show
evidence for standing slow-mode waves excited by small flarelike events \citep{mari08, erd08b}. These oscillations
have much smaller amplitudes than those of the SUMER oscillations that range from 12 to 353 km~s$^{-1}$
with an average value of 62$\pm$57 km~s$^{-1}$ \citep{wan05b}. \citet{mari08} found that the decay of the
EIS oscillations shows an intriguing temperature dependence that the decay appears to be
weaker seen in emission lines with the higher temperature of formation.  This behavior is inconsistent with the 
the damping mechanism for standing slow mode waves by thermal conduction and compressive viscosity
because they both cause the damping rate increasing with increasing temperature \citep{ofm02b}. 
To figure out whether the oscillations in higher- and lower-temperature regimes have different physical
properties and damping mechanisms first needs a reliable statistic study in the future. 

On the other hand, recently some studies argued that those quasi-periodic propagating disturbances 
in large fanlike loops observed with TRACE and EIT may not be the true waves but the signature of episodic (or
intermittent) heating events based on their association with tens of km~s$^{-1}$ outflows observed at 
the edges of active regions with Hinode/EIS \citep{sak07, har08, mci09, he10, de10}. It is very important to 
verify whether these small-amplitude Doppler-shift oscillations observed with EIS are longitudinal 
waves or quasi-periodic, heating events for better understanding their physical properties. 

\begin{figure}
\includegraphics[width=0.95\textwidth]{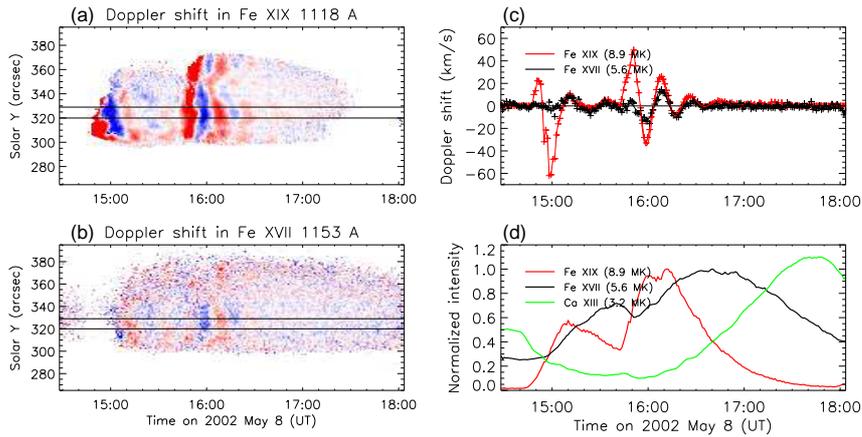}
\centering
\caption{Simultaneous observations of hot loop oscillations in two emission lines by SUMER. (a) Time series of
Doppler shift oscillation along the slit in Fe\,{\sc{xix}}. (b) The Doppler shift oscillation in Fe\,{\sc{xvii}}. 
(c) Time profiles of Doppler shift oscillations in Fe\,{\sc{xix}} and Fe\,{\sc{xvii}} averaged for a region marked 
in (a) and (b). The solid curves are the best fit to a damped sine function. (d) The corresponding average line 
intensities in Fe\,{\sc{xix}} $\lambda$1118.06, Fe\,{\sc{xvii}} $\lambda$1153.17 and Ca\,{\sc{xiii}} $\lambda$1133.76. }
\label{fgco} 
\end{figure}

In addition, the study of temporal dependence of wave properties on the temperature is another 
interesting topic, which is potentially applied in diagnostics of hydrodynamic processes related to impulsive heating 
of coronal loops.  Figure~\ref{fgco} shows an example for Doppler-shift oscillation events observed 
simultaneously in Fe\,{\sc{xix}} and Fe\,{\sc{xvii}} lines with SUMER. The intensity evolution in Fe\,{\sc{xix}}, 
Fe\,{\sc{xvii}} and Ca\,{\sc{xiii}} indicates a gradual cooling of the oscillating hot plasma 
in the loop (Fig.~\ref{fgco}d). The initial amplitudes of oscillations seen in  Fe\,{\sc{xix}} are
much larger than those in Fe\,{\sc{xvii}}, while later on they tend to become consistency (Fig.~\ref{fgco}c). 
The fitting to the second event (excluding the initial phase) shows that the period and decay time for
the oscillations in the two lines are almost the same. This implies that the oscillations are produced by 
the same plasma and appear to decay quicker than the loop cooling. In other words, the decay of oscillations 
seen in Fe\,{\sc{xix}} is real but not due to the cooling effect. Since the decay time (about 15 min) of 
oscillations in this event is shorter than the cooling time (about 30 min) of hot plasma seen in the emission 
from Fe\,{\sc{xix}} to Fe\,{\sc{xvii}} line, the oscillation has died out before the plasma cools to 
a lower temperature at which the oscillation is expected to have a longer decay time than that seen in 
Fe\,{\sc{xix}.} Otherwise the oscillation seen in Fe\,{\sc{xvii}} would last longer than in Fe\,{\sc{xix}}.

\subsection{Multiple harmonics}
Multiple harmonics of coronal loop kink-mode oscillations have been detected by TRACE and applied to
diagnose longitudinal structuring such as density stratification in coronal loops \citep[e.g.][]{ver04, and09}. 
For the slow mode several analytical studies showed that the longitudinal structuring or gravitational 
stratification can also modify the ratio between the period of the fundamental mode to that of the second harmonic 
strongly in certain conditions \citep{rob06, dia06, mce06b}. For example, this ratio strongly departs from the canonical 
value of 2 in coronal loops with high density contrast between footpoints and apex or in cooler loops. 
Therefore, the detection of multiple harmonics is important in development of coronal seismology. 

It was shown that Doppler-shift oscillations observed by SUMER in Fe\,{\sc{xix}} and those observed by Yohkoh/BCS 
in S\,{\sc{xv}} and Ca\,{\sc{xix}} all appear to be consistent with the fundamental slow mode 
\citep{wan03b, wan07, mari06}. Recently, \citet{sri10} stated that the evidence for multiple harmonics of slow 
acoustic oscillations was found in the nonflaring coronal loop with the Hinode/EIS data. 
However, the fact that the intensity signal with the period of the second harmonic is only 
detected at the loop apex but not at the footpoints does not completely agree with the theory. 

\citet{nak04b} and \citet{tsi04} simulated a high-temperature (20$-$30 MK) flaring loop using 
a 1D MHD model. They found that 
only the acoustic second standing harmonic is excited, regardless of the spatial position of heat deposition 
in the loop, and thus suggested that the quasi-periodic pulsations with periods in the range 
10$-$300 s, frequently observed in flaring coronal loops in the radio, visible light and X-ray bands, 
may be produced by the second harmonic slow mode. However, direct evidence has not yet been found 
in observation. \citet{nak06} found that the fundamental and second 
harmonic modes have quite distinct observational signatures in the loop's microwave images, which can 
be used to identify the mode in combination with analysis of the modulation frequency. In a different way, 
\citet{sel05} numerically studied the excitation of standing slow-mode waves in a hot ($\sim$5 MK) coronal 
loop by launching a hot pulse at different positions. They found that pulses close to a footpoint excite 
the fundamental mode, while pulses close the apex excite the second harmonic. In some cases, observed is
not a single mode but instead a packet of modes in which the fundamental and harmonic standing modes
make the largest contribution. Thus, the interpretation of the SUMER oscillations in terms of the 
fundamental slow mode implies the possible asymmetric (footpoint) excitation,
for which some pieces of evidence have been found in observations (see Sect.~\ref{scttrg}). 
In addition, it is noted that the wave decay rate of slow modes is larger for higher harmonics due
to viscous damping or thermal conduction damping \citep{por94, ofm00}. Thus, it is expected that
detection of the higher harmonics is more difficult than the fundamental mode, especially
in hot coronal loops.

\section{Summary}
The advance in observations and modelings of the standing slow mode waves have been reviewed in this paper.
These strongly damped Doppler-shift oscillations are commonly observed in flaring hot coronal loops in
active regions with different spectrometers such as the SUMER and BCS. The observed periodicities and
the phase relationship between the Doppler shift and intensity variations indicate that they are
fundamental standing slow-mode waves. The high excitation and recurrence rates of these oscillatory
events in coronal loops suggest that the standing slow mode waves are a nature of confined (or non-eruptive) 
small- or micro-flares, which are produced by impulsive heating in closed magnetic 
structure. The theoretical studies show that the understanding of quick excitation of the fundamental
mode needs at least the 2D MHD models with a curved geometry which allows the mode coupling to act, 
while the rapid damping can be well interpreted with the 1D nonlinear model.

Observations of the standing slow-mode oscillations in coronal loops are important for development of
coronal seismology. The measured period has been applied to determine the mean magnetic field
strength in coronal loops. The spatial and temporal relationship between Doppler shift and intensity
oscillations may be used to diagnose the cooling of the flare plasma in coronal loops.
In addition, the feature of phase propagation observed in Doppler shift oscillations by SUMER may have
a potential to be used for diagnostics of fine structure of coronal loops based on 3D MHD modeling.
 
The Doppler-shift oscillations with small amplitudes of several km s$^{-1}$ were observed with Hinode/EIS
in coronal loops of a temperature of 1$-$2 MK. These oscillations show a peculiar temperature-dependent
behavior that the damping appears to be weaker seen in the higher temperature emission lines. The
implication of this behavior is still unclear. So far, only the fundamental mode of slow mode oscillations
is detected in hot flaring loops, although the excitation of the second harmonic is possible as predicted
by numerical modeling with the symmetrical heating. The SDO/AIA with high spatio-temporal resolutions
and wide (1$-$10 MK) temperature coverage will provide better statistics in identification of different
(propagating or standing) oscillation modes and help elucidate the trigger sources and excitation
conditions.      

\begin{acknowledgements}
The author is grateful to Dr. Leon Ofman for his valuable comments in improving the manuscript. 
This work was supported by NASA grants NNX08AE44G and NNG06GI55G as well as NRL grant N00173-06-1-G033.
\end{acknowledgements}


\end{document}